# A Sampling Strategy in Efficient Potential Energy Surface Mapping for Predicting Atomic Diffusivity in Crystals by Machine Learning

Kazuaki Toyoura[*,1,2], Takeo Fujii[1], Kenta Kanamori[3], and Ichiro Takeuchi[2,3]

[1]*Department of Materials Science and Engineering, Kyoto University, Kyoto 606-8501, Japan*
[2]*RIKEN Center for Advanced Intelligence Project, Tokyo 103-0027, Japan*
[3]*Department of Computer Science, Nagoya Institute of Technology, Nagoya 466-8555, Japan*

**Abstract**

We propose a machine-learning-based (ML-based) method for efficiently predicting atomic diffusivity in crystals, in which the potential energy surface (PES) of a diffusion carrier is *partially* evaluated by first-principles calculations. To preferentially evaluate the *region of interest* governing the atomic diffusivity, a statistical PES model based on a *Gaussian process* (GP-PES) is constructed and updated iteratively from known information on already-computed potential energies (PEs). In the proposed method, all local energy minima (stable & metastable sites) and elementary processes of atomic diffusion (atomic jumps) are explored on the predictive mean of the GP-PES. The uncertainty of jump frequency in each elementary process is then estimated on the basis of the variance of the GP-PES. The acquisition function determining the next grid point to be computed is designed to reflect the impacts of the uncertainties of jump frequencies on the uncertainty of the macroscopic atomic diffusivity. A numerical solution of the master equation is here employed to readily estimate the atomic diffusivity, which enables us to design the acquisition function reflecting the centrality of each elementary process.

# I. INTRODUCTION

Atomic transport in solids is of significance in a wide range of phenomena concerning solid state physics and chemistry, metallurgy, and materials science. Theoretical approaches based on first-principles calculations are powerful techniques to clarify the microscopic picture of the atomic transport and to estimate the atomic diffusivity. There are two conventional atomic-scale simulations for estimating the atomic diffusivity, i.e., the molecular dynamics (MD) method [1-3] and the combination technique of the nudged elastic band (NEB) method [4,5] and the kinetic Monte Carlo (KMC) method [6,7]. The MD method reveals the time evolution of a system by numerically solving Newton's equations of motion with a fine time step. It has an advantage that the time evolution exactly follows classical Newtonian dynamics, but has a practical disadvantage, so-called *time-scale limitation* due to the fine time step of $\sim 10^{-15}$ s in the case of solid systems.

The latter technique of the NEB and KMC methods overcomes the time-scale limitation in the MD method. In the KMC method, a state transition and the elapsed time are stochastically determined at each KMC step according to the frequencies of all possible state transitions. Therefore, an atomic jump occurs at any step even if all possible atomic jumps at the step are rare events. In such a case, the elapsed time becomes longer at the step according to the low jump frequencies. An important task in this technique is preparing the list of all possible atomic jumps and their frequencies before the KMC simulations. The NEB method is often used for exploring possible elementary processes in a crystal, which finds valley lines on the potential energy surface (PES) of a diffusion carrier, so-called *minimum energy paths* (MEP). However, the NEB method requires prior knowledge on the initial and final states and the initial trajectory connecting the two states. The requirements are too demanding in complicated systems with low crystallographic symmetry, in which it is difficult to identify local minima on the PES and to specify an appropriate initial trajectory for each elementary process. The preparation of all possible elementary processes is therefore at risk of missing a key elementary process governing the atomic diffusivity, even though using our knowledge in solid state physics and chemistry.

An entire PES mapping for a diffusion carrier in a host crystal is a solid methodology for avoiding such a risk in the NEB method. The most straightforward method is that the potential energy (PE) of a diffusion carrier is computed at every grid point on a fine grid introduced in the host crystal. The PEs at the grid points only in the asymmetric unit should be computed, which are here called *irreducible grid points*. The number of the irreducible grid points drastically increases with lowering the crystallographic symmetry, proportional to the volume of the asymmetric unit. According to the Inorganic Crystal Structure Database (ICSD) [8], the asymmetric units of some oxides are larger than 1000 Å$^3$, meaning that the total number of the irreducible grid points is more than 10$^5$ even in the case of a somewhat rough grid (grid interval: ~ 0.2 Å). Therefore, the computational cost of the straightforward PES mapping is comparable to the MD method in some cases.

Recently, we have proposed two machine-leaning-based (ML-based) methods for efficient PES mapping [9,10], in which the PEs are computed preferentially from *region of interest* characterizing the atomic diffusivity. In both previous methods, the statistical PES model based on a *Gaussian process* (GP-PES) is constructed and updated iteratively using the already-computed PE information at earlier steps. On the basis of the GP-PES, the grid point with the highest probability that the grid point is located in the region of interest is selected as the next grid point to be computed. A notable advantage of the ML-based methods is *versatility* in the sense that these methods are applicable to different host crystals without any prior knowledge on the physical and/or chemical properties unlike the NEB method.

The major difference between the two previous methods is in the definition of the region of interest. In the first method [9], the region of interest is defined as the low-PE region which should include multiple local energy minima and elementary processes for the long-range migration. The first method however has the critical issue that the maximum reduction rate of the computational cost is governed by the ratio of the low PE region to the host crystal. In the case of the proton diffusion in barium zirconate with the cubic perovskite structure ($c$-BaZrO$_3$), the low PE region occupies ~ 20 % of the host crystal [9], meaning that the computational cost can be reduced by 80 % at most. In the second method [10], the region of interest is ultimately limited to overcome the efficiency limitation of the first method. The region of interest is defined as only a few dominant points, i.e., the global minimum point and the saddle point with the highest PE on the long-range migration path with the lowest potential barrier (called the *bottleneck point* on the *optimal path*, hereafter). Since the number of grid points characterizing the optimal path is much smaller than that in the low PE region, the second method is much more efficient than the first method. However, the second method cannot accurately evaluate the diffusivity and cannot clarify the detailed diffusion mechanism, because it focuses only on the potential barrier along the optimal path.

Thus, the first previous method is accurate but less efficient because the goal is to completely identify the PES in the low PE region containing a number of grid points. On the other hand, the second previous method is efficient but less accurate because the goal is to identify only a few dominant points characterizing the optimal path. In the present study, an ML-based method with both accuracy and efficiency has been proposed on the basis of the GP-PES model. In the rigorous manner, the jump frequencies of all elementary processes are required to accurately estimate the atomic diffusivity. However, a part of elementary processes have slight impact on the macroscopic diffusivity, whose jump frequencies are allowed to be roughly estimated. In the present proposed method, the acquisition function is designed to reflect the impact of the uncertainty of the jump frequency on the uncertainty of the diffusivity. A numerical solution of the master equation is here employed to quickly estimate the diffusivity and the impacts of the uncertainties of all jump frequencies, which enables the proposed method to be feasible. The efficiency and versatility of the proposed method are demonstrated on two examples, the isotropic and anisotropic proton diffusivities in $c$-BaZrO$_3$ and $t$-BaTiO$_3$ with the cubic and tetragonal perovskite structures.

## II. PROPOSED METHOD

In order to accurately and efficiently predict atomic diffusivity, computational techniques in physics and materials science are effectively combined with ML techniques such as a GP model and Bayesian optimization. The proposed ML-based method in the present study is the procedure with multiple steps as follows:

(0) Computing the PEs at initial grid points sampled at random (10 grid points in the present study).

(1) Constructing (updating) the GP-PES model based on the already-computed PEs.

(2) Identifying all elementary processes for atomic jumps on the predictive mean of the GP-PES.

(3) Estimating the jump frequency and the uncertainty in each elementary process.

(4) Calculating the acquisition value at each grid point based on the acquisition function reflecting the impact of the uncertainty of the jump frequency on the uncertainty of the atomic diffusivity.

(5) Selecting the next grid point with the highest acquisition value and computing the PE.

Procedures (1)-(5) are iteratively performed until the uncertainty of the diffusivity converges within a given accuracy. The details in each step are described in the following subsections. The major difference from the previously-proposed methods is in the sampling strategy for the next grid point to be computed, which is here designed suitable for converging the predicted atomic diffusivity with accuracy.

### A. PE computations

In the present study, the proton diffusion in $c$-BaZrO$_3$ and $t$-BaTiO$_3$ are taken as model systems for the application of the proposed method. The PE of a proton at each grid point in the host crystals was computed using first-principles calculations on the basis of the projector augmented wave (PAW) method as implemented in the VASP code [11-14]. The 5$s$, 5$p$, 6$s$ and 5$d$ orbitals for Ba, 3$p$, 4$s$ and 3$d$ orbitals for Ti, 4$s$, 4$p$, 5$s$ and 4$d$ orbitals for Zr, 2$s$ and 2$p$ orbitals for O, and 1$s$ orbital for H were treated as valence states in the PAW potentials. The generalized gradient approximation (GGA) parameterized by Perdew, Burke, and Ernzerhof was used for the exchange-correlation term [15]. The plane wave cut-off energies for the basis set and the augmentation charges were set to 400 eV and 605.4 eV, respectively. A supercell consisting of 3×3×3 unit cells were used with a 2×2×2 mesh for the $k$-point sampling, in which the atomic positions were optimized with fixing the proton at the grid point and the farthest cation from the proton. The numbers of irreducible grid points in $c$-BaZrO$_3$ and $t$-BaTiO$_3$ are 286 and 720, respectively (See Figs. 3(a) and 7(a)).

**B. Statistical PES model based on Gaussian process**

In the present study, a statistical PES model based on a GP [16,17] is employed as in our previous studies [9,10]. Using the GP model, the predictive distributions of the potential energy $E_i$ at each irreducible grid point $i$ ($i = 1, …, n_{asym}$) is completely characterized by the mean function $\mu(\mathbf{x})$ and the covariance function $k(\mathbf{x}, \mathbf{x}')$, where $\mathbf{x}$ is the three-dimensional coordinates of the grid point. $k(\mathbf{x}, \mathbf{x}')$ is the so-called *kernel function*, interpreted as the similarity between the coordinates $\mathbf{x}$ and $\mathbf{x}'$. One of the most commonly used kernel functions is the RBF kernel, given by

$$k(\mathbf{x}, \mathbf{x}') = \sigma_f^2 \exp(-\|\mathbf{x} - \mathbf{x}'\|^2 / 2l^2), \qquad (1)$$

where $\sigma_f$ and $l$ ($> 0$) are tuning parameters, which are determined by maximizing the marginal likelihood at each iteration. On the other hand, the constant function is commonly used for the mean function $\mu(\mathbf{x})$, i.e., $\mu(\mathbf{x}) = \mu_0$.

The GP-PES model based on the already-computed PEs at $m$ grid points provides the predictive distribution of the entire PES in the form of a normal distribution $N[\mu_m(\mathbf{x}), \sigma_m^2(\mathbf{x})]$ at each grid point $\mathbf{x}$. The predictive mean and variance are given by

$$\mu_m(\mathbf{x}) = \mu(\mathbf{x}) + k(\mathbf{x})^T \mathbf{K}^{-1}(\mathbf{E} - \boldsymbol{\mu}), \qquad (2)$$

$$\sigma_m^2(\mathbf{x}) = k(\mathbf{x}, \mathbf{x}) - k(\mathbf{x})^T \mathbf{K}^{-1} k(\mathbf{x}), \qquad (3)$$

where $k(\mathbf{x}) = [k(\mathbf{x}, \mathbf{x}_1), …, k(\mathbf{x}, \mathbf{x}_m)]^T$, $\mathbf{E} = [E_1, …, E_m]^T$, $\boldsymbol{\mu} = [\mu(\mathbf{x}_1), …, \mu(\mathbf{x}_m)]^T$, and $\mathbf{K}$ is the so-called *kernel matrix* defined as

$$\mathbf{K} = \begin{bmatrix} k(\mathbf{x}_1, \mathbf{x}_1) & \cdots & k(\mathbf{x}_1, \mathbf{x}_m) \\ \vdots & \ddots & \vdots \\ k(\mathbf{x}_m, \mathbf{x}_1) & \cdots & k(\mathbf{x}_m, \mathbf{x}_m) \end{bmatrix}. \tag{4}$$

Thus, the GP-PES provides not only the predictive PES (mean) but also the uncertainty (variance), which enables us to make a sampling strategy for the next grid point to be computed. The conventional strategies are based on the probability of improvement (PI) and the expected improvement (EI), which are often used in the literature of Bayesian optimization including one of our previous studies [9,18-25].

**C. Elementary processes and jump frequencies**

All elementary processes in the unit cell are explored on the predictive mean of GP-PES with two steps, i.e., identifying all local energy minima (sites) and finding elementary processes (atomic jumps) connecting adjacent sites.

At the first step, all local minima on the predictive mean of GP-PES are identified by comparing the PE at every grid point with the PEs at all adjacent grid points. Specifically, the grid point $i$ is a local energy minimum if satisfying the following condition,

$$\hat{E}_i < \hat{E}_j \quad (^\forall j \in A_i), \tag{5}$$

where $\hat{E}_i$ is the predicted PE at the grid point $i$, and $A_i$ is the set of all grid points adjacent to the grid point $i$.

At the second step, all valley lines connecting adjacent sites are explored on the predicted PES. $G$ and $S$ are defined as the sets of the indexes of all grid points and the identified local minima (sites) in the unit cell, respectively,

$$G = \{1, 2, \ldots, n_{\text{grid}}\}, \tag{6}$$

$$S = \{s_1, s_2, \ldots, s_{n_{\text{site}}}\}, \tag{7}$$

where $n_{\text{grid}}$ and $n_{\text{site}}$ are the numbers of all grid points and sites in the unit cell, respectively, and $s_i$ is the grid point indexes of all sites. Here a *basin* is defined around each site, which is separated by several saddle surfaces as shown in Fig. 1. $B_i$ is defined as the set of grid point indexes in the basin around a site $s_i$. As the initial operations before exploring elementary processes for atomic jumps, the grid point index of each site $i$ is extracted from the set $G$ ($G \leftarrow G \setminus \{s_i\}$) and added into the basin set $B_i$ ($B_i \leftarrow B_i \cup \{s_i\}$). Then, every grid point $k$ is extracted from the set $G$ one by one in ascending order of the predicted PE ($G \leftarrow G \setminus \{k\}$). If the extracted grid point $k$ is an adjacent point to a grid point in the set of $B_i$ ($k \in A_j$ ($\exists j \in B_i$)), $k$ is added into the basin set $B_i$ ($B_i \leftarrow B_i \cup \{k\}$). When a grid point is added into two basin sets $B_i$ and $B_j$ at the same time, an elementary process exists between the two sites where the added grid point is the saddle point. The trajectory of the elementary process is identified by tracing the adjacent grid point from the saddle point in the steepest descent direction. Note that the adjacent basins are separated by not only the saddle point but the saddle surface consisting of several grid points. Therefore, a common grid point is added into two basin sets ($B_i$ and $B_j$) only when the product set is empty ($B_i \cap B_j = \emptyset$). This enables us to explore several elementary processes for atomic jumps from a single site.

The jump frequency from site $i$ to site $j$, $\Gamma_{ij}$, is estimated on the basis of the transition state theory (TST) [26-28] as follows:

$$\Gamma_{ij} = \Gamma_0 \exp(-\Delta E_{ij}^{\text{mig}}/k_\text{B}T), \tag{8}$$

where $\Gamma_0$ is the vibrational prefactor, $\Delta E_{ij}^{\text{mig}}$ is the potential barrier, $k_\text{B}$ is the Boltzmann constant, and $T$ is the temperature. In the present study, $\Gamma_0$ for all paths were set to a typical value for ionic jumps in crystals, i.e., 10 THz [29-33].

**D. Numerical solution of diffusivity by master equation**

In the proposed method, the sampling strategy is based on atomic diffusivity, which therefore have to be estimated frequently during the PES mapping. The conventional KMC method is cumbrous for the frequent estimation of atomic diffusivity due to setting the simulation conditions, i.e., the numbers of steps and trials. Instead of the KMC method, a numerical solution of the master equation [34,35] is employed in the present study, which finally results in a simple eigenvalue problem of a matrix.

Under the independent-particle approximation, the master equation corresponding to the balance of the existence probability of a single particle at each site $i$, $p_i$, is given by

$$\frac{\partial p_i(t)}{\partial t} = \sum_j [\Gamma_{ji} p_j(t) - \Gamma_{ij} p_i(t)], \tag{9}$$

where $t$ is the time, and $\Gamma_{ij}$ is zero in the case of no elementary process between sites $i$ and $j$. The first and second terms in the brackets on the right side are the inflow and outflow of the existence probability, respectively. In an $N$-site system ($N$: number of sites), the jump frequency matrix $\Gamma$ is defined as the negative of the $N \times N$ Laplacian matrix for a weighted directed graph, in which the off-diagonal elements are $\Gamma_{ij}$ and the diagonal elements are $-\sum_{j \neq i} \Gamma_{ij}$. Using the matrix $\Gamma$, Eq. (9) can be expressed simply as

$$\frac{\partial \mathbf{p}}{\partial t} = \Gamma^{\mathrm{T}} \mathbf{p}, \tag{10}$$

where $\mathbf{p}$ is the vector of the existence probabilities of the single particle at all $N$ sites, $\mathbf{p} = [p_1(t), ..., p_N(t)]^{\mathrm{T}}$. The solution of the master equation is expressed using a given initial condition $\mathbf{p_0}$ at $t = 0$,

$$\mathbf{p} = \exp(\Gamma^{\mathrm{T}} t) \mathbf{p_0}. \tag{11}$$

The diffusivity can thereby be evaluated by estimating the mean square displacement of the particle from the time dependence of the existence probability.

In a crystal, the number of sites $N$ is extremely large in general, but the dimensions of the matrix $\Gamma$ and the vector $\mathbf{p}$ can be reduced drastically by exploiting the translational symmetry. The existence probability $p_i$ is hereafter redefined as a function of the position $\mathbf{r}$ in addition to the time $t$, $p_i(\mathbf{r}, t)$, where $i$ is the site index in the unit cell ($i = 1, ..., n_{\mathrm{site}}$) and $\mathbf{r}$ is defined in the global coordinate system. The master equation of Eq. (9) is rewritten as

$$\frac{\partial p_i(\mathbf{r},t)}{\partial t} = \sum_{j,\alpha} \left[ \Gamma_{ji}^{\alpha} p_j(\mathbf{r}+\mathbf{s}_{ij}^{\alpha},t) - \Gamma_{ij}^{\alpha} p_i(\mathbf{r},t) \right]. \tag{12}$$

The summation on the right side is made for all adjacent sites beyond the periodic boundaries around site $i$ in the focused unit cell. When site $i$ has several adjacent sites $j$ in different unit cells, they are distinguished by the unit cell index $\alpha$. $\mathbf{s}_{ij}^{\alpha}$ is the jump vector from site $i$ in the focused unit cell to site $j$ in unit cell $\alpha$.

The reduced master equation (Eq. (12)) can easily be solved in the reciprocal space rather than in the real space. With the Fourier transform, $p_i(\mathbf{r}, t)$ is transformed into $P_i(\mathbf{Q}, t)$,

$$P_i(\mathbf{Q}, t) = \int p_i(\mathbf{r}, t)\exp(i\mathbf{Q}\mathbf{r})d\mathbf{r}, \tag{13}$$

where $\mathbf{Q}$ is the Fourier variable corresponding to position $\mathbf{r}$. Eq. (12) is also transformed as follows:

$$\frac{\partial P_i(\mathbf{Q},t)}{\partial t} = \sum_{j,\alpha}\left[\Gamma_{ji}^{\alpha}P_j(\mathbf{Q},t)\exp(-i\mathbf{Q}\mathbf{s}_{ij}^{\alpha}) - \Gamma_{ij}^{\alpha}P_i(\mathbf{Q},t)\right]$$

$$= \sum_j\left[\sum_{\alpha}\Gamma_{ji}^{\alpha}\exp(-i\mathbf{Q}\mathbf{s}_{ij}^{\alpha})\right]P_j(\mathbf{Q},t) - \left[\sum_{j,\alpha}\Gamma_{ij}^{\alpha}\right]P_i(\mathbf{Q},t). \tag{14}$$

When the jump matrix $\mathbf{\Lambda}$ is defined as an $n_{\text{site}} \times n_{\text{site}}$ matrix with the elements $\Lambda_{ij}$,

$$\Lambda_{ij} = \sum_{\alpha}\Gamma_{ij}^{\alpha}\exp(i\mathbf{Q}\mathbf{s}_{ij}^{\alpha}) - \delta_{ij}\sum_{j',\alpha}\Gamma_{ij'}^{\alpha} \quad (\delta_{ij}\text{: Kronecker delta}), \tag{15}$$

the master equation in the reciprocal space is simply expressed as

$$\frac{\partial \mathbf{P}}{\partial t} = \mathbf{\Lambda}^{\mathrm{T}}\mathbf{P}, \tag{16}$$

which is the similar matrix expression to that in the real space (Eq. (10)). $\mathbf{P}$ is the vector of the existence probabilities at all $n_{\text{site}}$ sites in the reciprocal space, $\mathbf{P} = [P_1(\mathbf{Q}, t), ..., P_{n_{\text{site}}}(\mathbf{Q}, t)]^{\mathrm{T}}$. Thus, the dimensions of the existence-probability vector and the jump-frequency matrix are reduced from $N$ and $N \times N$" to "$n_{\text{site}}$ and $n_{\text{site}} \times n_{\text{site}}$" by exploiting the translational symmetry.

All the eigenvalues of $\mathbf{\Lambda}^{\mathrm{T}}$ are negative real numbers for any $\mathbf{Q}$ vector with a non-zero magnitude. When $\mathbf{\Lambda}^{\mathrm{T}}$ is expressed as $\mathbf{XYX}^{-1}$ using the eigenvalue diagonal matrix $\mathbf{Y}$ (eigenvalues: $\lambda_i$) and the transformation matrix $\mathbf{X}$ for the diagonalization, the solution is transformed as follows:

$$\mathbf{P} = \exp(\mathbf{\Lambda}^{\mathrm{T}}t)\mathbf{P}_0 = \exp(\mathbf{XYX}^{-1}t)\mathbf{P}_0$$

$$= \mathbf{X}\exp(\mathbf{Y}t)\mathbf{X}^{-1}\mathbf{P}_0 = \mathbf{X}\begin{pmatrix} e^{\lambda_1 t} & & 0 \\ & \ddots & \\ 0 & & e^{\lambda_n t} \end{pmatrix}\mathbf{X}^{-1}\mathbf{P}_0. \tag{17}$$

This indicates that the existence probability distribution of a single particle in the reciprocal space $P(\mathbf{Q}, t) = \sum_i P_i(\mathbf{Q}, t)$ is expressed as the summation of multiple exponential terms.

To connect the solution of the master equation in the reciprocal space with the diffusion coefficient tensor $\mathbf{D}$, the following Fick's second law defined in continuous space is solved.

$$\frac{\partial p(\mathbf{r},t)}{\partial t} = \nabla[\mathbf{D}\nabla p(\mathbf{r}, t)], \tag{18}$$

where $p(\mathbf{r}, t)$ is the existence probability distribution of a single particle at time $t$. With the Fourier transform, Eq. (18) and the initial condition of $p(\mathbf{r}, 0) = \delta(\mathbf{r})$ are transformed into

$$\frac{\partial P(\mathbf{Q},t)}{\partial t} = -\sum_{m,n=x,y,z} D_{mn} Q_m Q_n P(\mathbf{Q}, t), \tag{19}$$

$$P(\mathbf{Q}, 0) = 1, \tag{20}$$

where $D_{mn}$ is the elements of the diffusion coefficient tensor $\mathbf{D}$. The solution of Eq. (19) is expressed as a single exponential term,

$$P(\mathbf{Q}, t) = \exp\left[\left(-\sum_{m,n=x,y,z} D_{mn} Q_m Q_n\right)t\right], \tag{21}$$

Considering the time and spatial scales of atomic diffusion ($t \to \infty$ and $|\mathbf{Q}| \to 0$), Eq. (17) having multiple exponential terms coincides with Eq. (21) expressed by a single exponential term, that is, the maximum eigenvalue $\lambda_1$ is equal to $-\sum_{m,n=x,y,z} D_{mn} Q_m Q_n$. Consequently, all the elements $D_{mn}$ can be obtained by solving the eigenvalue problems for properly-selected $\mathbf{Q}$ vectors with a small magnitude relative to the scale of Brillouin Zone. The diffusion coefficient tensor $\mathbf{D}$ is a real symmetric matrix, which can be diagonalized to transform the Cartesian coordinate system to the principal-axis coordinate system, $\mathbf{D}^{\text{diag}}$, if necessary.

**E. Acquisition function**

The acquisition function in the present study is designed to reflect the impact of the uncertainty for each jump frequency on the atomic diffusivity. Figure 2(a) shows the schematic diagram of the GP-PES along the trajectory of an elementary process from the initial point $i$ to the final point $j$ through the saddle point $s$. The uncertainty of the jump frequency $\Gamma_{ij}$ involves two major factors concerning the uncertainties of the initial and saddle points. Based on the lower confidence bound (LCB), the PE in the initial state can become lower from $\hat{E}_i$ to the lowest PE between the initial and saddle points on the LCB, $\hat{E}_{i'}^{\text{LCB}}$. Similarly, the PE in the saddle-point state can become higher from $\hat{E}_s$ to the highest PE between the initial and final points on the upper confidence bound (UCB), $\hat{E}_{s'}^{\text{UCB}}$. These two uncertainties lead to the uncertainty of the jump frequency in the elementary process. The 95% confidence interval was employed for the LCB and UCB in the present study. Note that $\hat{E}_{s'}^{\text{UCB}} - \hat{E}_{i'}^{\text{LCB}}$ does not exactly correspond to the 95 % confidence interval of the potential barrier.

The above two uncertainties are limited only to those along the trajectory of the elementary process found on the predictive mean of the GP-PES. The uncertainty of the trajectory itself should also be taken into consideration, particularly around the initial and saddle points. In the present study, the most likely grid points to be the initial and saddle points are explored around the initial and saddle points $i$ and $s$ on the predictive mean of the GP-PES (Fig. 2(b)). Specifically, on the LCB, the grid point with the lowest PE in all adjacent grid points to the initial point $i$ is regarded as the initial point candidate (grid points $i''$), i.e., $i'' = \underset{k \in A_i}{\mathrm{argmin}}\, \hat{E}_k^{\text{LCB}}$. Similarly, the saddle point candidate is the grid point $s''$ satisfying the following equation, $s'' = \underset{k \in A_s}{\mathrm{argmin}}\, \hat{E}_k^{\text{LCB}}$. Note that the grid points on the trajectory of the elementary process are here excluded in the sets of adjacent grid points $A_i$ and $A_s$.

In the present study, the four uncertainties of the jump frequency for an elementary process are individually treated as follows:

$$\Gamma_{ij}^{i'} = \Gamma_0 \exp\bigl(-(\hat{E}_s - \hat{E}_{i'}^{\text{LCB}})/k_{\text{B}}T\bigr), \tag{22}$$

$$\Gamma_{ij}^{s'} = \Gamma_0 \exp\bigl(-(\hat{E}_{s'}^{\text{UCB}} - \hat{E}_i)/k_{\text{B}}T\bigr), \tag{23}$$

$$\Gamma_{ij}^{i''} = \Gamma_0 \exp\bigl(-(\hat{E}_s - \hat{E}_{i''}^{\text{LCB}})/k_{\text{B}}T\bigr), \tag{24}$$

$$\Gamma_{ij}^{s''} = \Gamma_0 \exp\bigl(-(\hat{E}_{s''}^{\text{LCB}} - \hat{E}_i)/k_{\text{B}}T\bigr). \tag{25}$$

The acquisition values at the four grid points *i'*, *s'*, *i''*, and *s''* are defined as the change ratio in the estimated diffusivity,

$$a(\mathbf{x}_k, k \in \{i', s', i'', s''\}) = \max\left(\frac{D_1^k}{D_1}, \frac{D_2^k}{D_2}, \frac{D_3^k}{D_3}, \frac{D_1}{D_1^k}, \frac{D_2}{D_2^k}, \frac{D_3}{D_3^k}\right) - 1, \qquad (26)$$

where $D_n$ ($n$ = 1, 2, 3) are the three elements in the diagonalized diffusion coefficient tensor $\mathbf{D}^{\text{diag}}$ estimated on the predictive mean of the GP-PES, and $D_n^k$ are those in the case that $\Gamma_{ij}$ is replaced by $\Gamma_{ij}^k$. All jump frequencies for the equivalent elementary processes in the unit cell are replaced by $\Gamma_{ij}^k$ to estimate the $D_n^k$. Consequently, the acquisition function $a(\mathbf{x})$ has a finite value only when $\mathbf{x}$ corresponds to grid points *i'*, *i''*, *s'*, or *s''* for any elementary process, and zero otherwise.

The stopping criterion for the PES mapping is based on the acquisition function in the present study. Specifically, the grid point sampling is stopped when all the acquisition values become less than a given threshold, e.g., $10^{-5}$.

## III. RESULTS & DISCUSSION

### A. Proton PES in BaZrO$_3$

The proposed method in the present study is first applied to the isotropic proton diffusion in *c*-BaZrO$_3$, which is characterized only by a single diffusion coefficient. Figure 3 shows all irreducible grid points (286 points) in the asymmetric unit and the entire PES of a single proton obtained by exhaustive PE computations at all irreducible grid points by first-principles calculations. The yellow surface is the PE isosurface (isosurface level: 0.3 eV vs. the global minimum point), indicating the low PE regions. The pale red and blue spheres denote the global minimum points (min 1) and the trajectories of proton jumps connecting the global minimum points, respectively. There are eight global minimum points around a single oxide ion, all of which are equivalent crystallographically in this crystal. The closest two global minimum points are connected by the trajectory through a saddle point (saddle 1) with a quite low PE (0.01 eV), which can be regarded as *oscillatory proton transfer*. The other two trajectories through saddle 2 and saddle 3 correspond to *proton rotation* around an oxide ion and *proton hopping* between two adjacent oxide ions, respectively. The calculated potential barriers of proton rotation and hopping are 0.19 and 0.29 eV, respectively.

In the literature, the atomic-scale picture of proton diffusion in *c*-BaZrO$_3$ has intensively been investigated in a first-principles manner [36-42]. The first-principles MD simulations [36] already revealed the rotation and hopping mechanism of proton diffusion in the crystal. In our previous study based on the NEB method [41], the calculated potential barriers for the proton rotation and hopping in undoped BaZrO$_3$ are 0.17 eV and 0.25 eV, respectively. The slight differences in potential barrier between the present and previous studies are due to the relatively coarse grid for the PES mapping in the present study. Actually, the calculated potential barriers on the PES with a finer grid (40×40×40 in the unit cell) [9,10] are 0.18 and 0.25 eV, almost equal to the potential barriers by the NEB method. The presence or absence of the oscillatory proton transfer also depends on the computational conditions. According to our recent study [41], the global minimum point in a larger supercell consisting of 4×4×4 unit cells is located at the saddle point of the oscillatory proton transfer (saddle 1), meaning that the supercell of 3×3×3 unit cells employed in the present study is rather small for describing the local structural relaxation around a single proton and that such oscillatory proton transfer does not occur in the case of dilute protons. Nevertheless, the proton PES with three types of proton jumps is here regarded as a true PES for the performance test of the proposed method, which is more difficult to predict than the simple PES with only proton rotation and hopping. The estimated diffusion coefficient of protons on the true PES is $4.3 \times 10^{-5}$ cm$^2$/s at 1000 K.

Figure 4 shows a typical profile of the predicted diffusion coefficient of protons at 1000 K as a function of the number of PE computations, which includes 10 PE computations at the initial grid points randomly sampled. The red line shows the diffusion coefficient on the predictive mean of the GP-PES, and the pale red region denotes the uncertainty estimated from the 95 % confidence interval of the GP-PES. The uncertainty is two order of magnitude at the beginning of the preferential sampling, but it gradually decreases, finally to be negligible after 30 PE computations. The converged diffusion coefficient is $4.3 \times 10^{-5}$ cm$^2$/s, exactly equal to the true one estimated using the entire PES. Figure 5 shows the sampled grid points and the PE isosurface (isosurface level: 0.3 eV) on the predictive mean of the GP-PES after 10, 15, 20, 25, 30, and 35 PE computations. After the 10 PE computations at the initial grid points (white spheres), the predicted PES is different from that on the true entire PES (Fig. 3(b)). During the preferential sampling, the grid points at the vicinity of the global minimum point and the three saddle points are preferentially sampled (black spheres), and the predicted PES becomes similar to the true one gradually. The single type of the global minimum points and the trajectories of elementary processes on the predictive mean of the GP-PES at each iteration are also shown in the figure. At the beginning, the position and the number of the global minimum points are incorrect, but they all are correctly identified after 25 PE computations. The predicted trajectories of elementary processes for proton jumps also converge after 30 PE computation, although the rotational trajectory is slightly different from the true one (Fig. 3(b)).

Figures S1(a) in the Supplementary Materials shows the profiles of the predicted diffusion coefficients in ten trials with different initial grid points. For comparison, those by the previous method [10] to preferentially sample only the global minimum and bottleneck points are also shown in Fig. S1(b). The present method requires 33–42 PE computations until the uncertainty of the predicted diffusion coefficient is negligible, which are comparable to the numbers of PE computations in the previous method (32–47 PE computations). Of particular note is that all predicted diffusion coefficients converge to the true value 4.3 cm$^2$/s in the present method, in contrast to the scattering of the final values in the previous method (2.3–4.3 cm$^2$/s). This excellent performance of the present method reflects the sampling strategy to converge the diffusion coefficient directly. In the previous method, the next grid point is sampled as the uncertainty of the potential barrier along the optimal path decreases, meaning that only the global minimum point and the saddle point of the proton hopping (bottleneck point) are sampled preferentially. Figure 6 shows the number of times that the global minimum and the three saddle points were sampled in the ten trials. The present method never fail to sample the four dominant points, while the previous method sometimes fail to sample some saddle points except for the global minimum and bottleneck points, leading to the misprediction.

**B. Sampling profiles in *t*-BaTiO$_3$**

The next example is the anisotropic proton diffusion in *t*-BaTiO$_3$, characterized by two independent diffusion coefficients in the *ab*-plane and along the *c*-axis. Figure 7(a) shows the irreducible grid points in the asymmetric unit (720 points), which are more than those in *c*-BaZrO$_3$ reflecting the lower crystallographic symmetry of the tetragonal perovskite structure. In the converged crystal structure after the structural optimization, the Ti ion is displaced by ~ 0.1 Å along the *c*-axis with reference to the anion sublattice. The symmetry reduction results in several types of local energy minima (proton sites) in *t*-BaTiO$_3$. Figure 7(b) shows six types of local minimum points on the entire PES of a single proton in the crystal obtained by exhaustive PE computations at all irreducible grid points by first-principles calculations (min 1–6). The blue spheres denote the global minimum points (min 1), located around O2 ions. The other two types of local minima around O2 ions (min 2 & 3) have a higher PE than the global minima by ~ 0.25 eV. On the other hand, the PEs of the local minima around O1 ions (min 4–6) are relatively high, in the range of 0.29–0.36 eV. The yellow surface in the figure is the PE isosurface (isosurface level: 0.5 eV vs. the global minima), showing the proton migration pathways. According to this figure, protons can migrate over a long range by way of only O1 ions in the *ab*-plane, while the long-range migration along the *c*-axis requires going through both O1 and O2 ions. As a result, the potential barrier of the optimal path along the *c*-axis is higher than that in the *ab*-plane, 0.48 eV vs. 0.32 eV.

Figure 8 shows the profiles of the predicted diffusion coefficients of protons at 1000 K in the *ab*-plane and along the *c*-axis as a function of the number of PE computations. Reflecting a lot of irreducible grid points and the complicated migration pathways, many PE computations (~ 150 grid points) are required in this system. The converged diffusion coefficients are $7.1 \times 10^{-5}$ cm$^2$/s in the *ab*-plane and $4.3 \times 10^{-5}$ cm$^2$/s along the *c*-axis, which are in reasonable agreement with the true diffusion coefficients estimated on the true PES ($7.5 \times 10^{-5}$ cm$^2$/s in the *ab*-plane and $4.3 \times 10^{-5}$ cm$^2$/s along the *c*-axis). Figure S2 in the Supplementary Materials shows the profiles of the predicted diffusion coefficients in ten trials with different initial grid points sampled at random. For comparison, those by the previous method [10], which preferentially samples only the global minimum and bottleneck points, are also shown in the figure. Although the present method requires more PE computations (150–190 computations) than the previous method (90–140 computations), the converged diffusion coefficients in the present method are closer to the true diffusion coefficients than those in the previous method. The root mean square errors (RMSEs) of the predicted diffusion coefficients are $0.4 \times 10^{-5}$ cm$^2$/s in the *ab*-plane and $0.2 \times 10^{-5}$ cm$^2$/s along the *c*-axis in the present method. On the other hand, the RMSEs are $0.9 \times 10^{-5}$ cm$^2$/s in the *ab*-plane and $0.8 \times 10^{-5}$ cm$^2$/s along the *c*-axis in the previous method, less accurate than the present method.

The difference in the accuracy between the two methods is attributed to the sampled grid points for PE computations. Figure 9 shows the numbers of times that the true local minima (min 1–6) and the true saddle points (saddle 1–14) were sampled in the ten trials. In the present method, 5 local minima and 10 saddle points were sampled in all trials without fail, while only 2 local minima and 3 saddle points were robustly sampled in the previous method. Focusing on the three dominant points, both methods almost certainly sampled the global minimum point and the two bottleneck points, all of which are indispensable to predict the diffusivity and the potential barriers. A few failures of sampling the bottleneck point on the optimal path in the *ab*-plane (bottleneck 1) is due to the alternate pathway through another saddle point (saddle 6). The PE at the saddle point along the alternate pathway (0.33 eV) is comparable to that at the bottleneck point (0.32 eV), leading to the reasonable prediction of the diffusivity even without the bottleneck point. Actually, the predicted potential barriers accurately converge to the true ones in both methods, as shown in Fig. S3 in the Supplementary Materials.

Thus, the performances of the present and previous methods mean the trade-off relation between the accuracy and the computational cost, suggesting that the two methods should be employed depending on the research purpose. The present method is better for understanding the detailed diffusion mechanism and estimating the accurate diffusivity, even if it requires higher computational cost. On the other hand, the previous method should be used for roughly estimating the diffusivity with as low computational cost as possible, e.g., screening a lot of candidates for materials exploration.

**IV. CONCLUSIONS**

In the present study, a ML-based method for efficient PES mapping of a diffusion carrier in a host crystal was proposed, in which the *region of interest* governing the atomic diffusivity are preferentially evaluated. During the PES mapping, a GP-PES model is constructed and updated iteratively from known information on already-computed PEs. In the proposed method, all local energy minima and elementary processes of atomic diffusion are explored on the predictive mean of the GP-PES. The uncertainty of jump frequency in each elementary process is then estimated on the basis of the LCB and UCB of the GP-PES. The acquisition function determining the next grid point to be computed is designed to reflect the impacts of the uncertainties of jump frequencies on the atomic diffusivity. The numerical solution of the master equation is here employed to frequently estimate the atomic diffusivity, which enables us to design the acquisition function reflecting the centrality of each elementary process.

The proposed method in the present study was applied to the isotropic and anisotropic proton diffusion in *c*-$BaZrO_3$ and *t*-$BaTiO_3$. In the case of *c*-$BaZrO_3$, the present method can accurately and robustly estimate the diffusion coefficient of protons with ~ 35 PE computations for 286 irreducible grid points. The required number of PE computations is comparable to that at the previous method focusing only on the two dominant points, i.e., the global minimum and bottleneck points. The present method exhibits higher accuracy of the diffusivity prediction, which is the advantage over the previous method. Even in the case of *t*-$BaTiO_3$ with more complicated PES, the present method still has the advantage of the accurate diffusivity prediction, but requires more PE computations than the previous method. This is due to the difference in the number of grid points to be explored between the two methods. In the previous method, only a type of the global minima and two types of bottleneck points are explored out of 6 local minima and 14 saddle points, while more dominant points are required to estimate the diffusion coefficient accurately. Actually, the present method sampled 5 local minima and 10 saddle points in all 10 trials without fail. The difference in performance between the present and previous methods indicates the trade-off relation between the accuracy and the computational cost, suggesting that the two methods should be employed depending on the research purpose.

**ACKNOWLEDGMENT**

This work was partially supported by grants from the Japanese Ministry of Education, Culture, Sports, Science and Technology to K.T. (17H04948, 19H05787), and I.T. (17H00758), and RIKEN Center for Advanced Intelligence Project to K.T. and I.T.

**Figure captions**

FIG. 1. (Color online) A basin surrounded by the four saddle surfaces (yellow lines) on a synthetic 2D-PES.

FIG. 2. (Color online) (a) Schematic diagram of the GP-PES along the trajectory of an elementary process from the initial point $i$ to the final point $j$ through the saddle point $s$. The red line and the pale blue area denote the predictive mean and the confidence interval, respectively. (b) Candidates of points $i''$ and $s''$ on the predictive mean of the GP-PES (two-dimensional PES), which are adjacent to the initial and saddle points, respectively. Note that the grid points on the trajectory of the elementary process are excluded from the $i''$ and $s''$ candidates.

FIG. 3. (Color online) (a) All irreducible grid points in the asymmetric unit of the $c$-$BaZrO_3$ crystal. (b) The global minimum points (pale red spheres) and the trajectories (pale blue spheres) of three types of elementary processes for proton jumps in $c$-$BaZrO_3$ (*oscillatory proton transfer*, *proton rotation*, and *proton hopping*). The yellow surface denotes the PE isosurface (isosurface level: 0.3 eV vs. the global minimum point).

FIG. 4. (Color online) The predicted diffusion coefficient of protons at 1000 K in $c$-$BaZrO_3$ with the uncertainty as a function of the number of PE computations. The first ten PE computations corresponds to the initial points sampled at random. The red line is the diffusion coefficient on the predictive mean of the GP-PES. The pale red region denotes the uncertainty of the predicted diffusion coefficient estimated from the 95 % confidence interval of the GP-PES.

FIG. 5. (Color online) The sampled grid points (white and black spheres), local minima (pale red spheres) and trajectories of elementary processes for proton jumps (pale blue spheres) on the predictive mean of the GP-PES after 10, 15, 20, 25, 30, and 35 PE computations. The white and black spheres correspond to the initial random sampling and the subsequent preferential sampling, respectively. The yellow surface denotes the PE isosurface on the predictive mean of the GP-PES (isosurface level: 0.3 eV vs. the global minimum point).

FIG. 6. (Color online) The comparison between the present method and the previous method [10] in the number of times that the global minimum and three saddle points in $c$-BaZrO$_3$ were sampled in the ten trials.

FIG. 7. (Color online) (a) All irreducible grid points in the asymmetric unit of the $t$-BaTiO$_3$ crystal. (b) The six local energy minima (small spheres bonding to single oxide ions) in $t$-BaTiO$_3$. The relative potential energies at the local energy minima are shown in parentheses. The yellow surface denotes the PE isosurface (isosurface level: 0.5 eV vs. the global minimum point).

FIG. 8. (Color online) The predicted diffusion coefficients of protons at 1000 K in $t$-BaTiO$_3$ with the uncertainty as a function of the number of PE computations. The first ten PE computations correspond to the initial points sampled at random. The red and blue lines are the diffusion coefficients in the $ab$-plane and along the $c$-axis on the predictive mean of the GP-PES, respectively. The pale red and blue regions denote the uncertainties of the predicted diffusion coefficients estimated from the 95 % confidence interval of the GP-PES.

FIG. 9. (Color online) The comparison between the present method and the previous method [10] in the number of times that the true local minima and saddle points in $t$-BaTiO$_3$ were sampled in the ten trials.

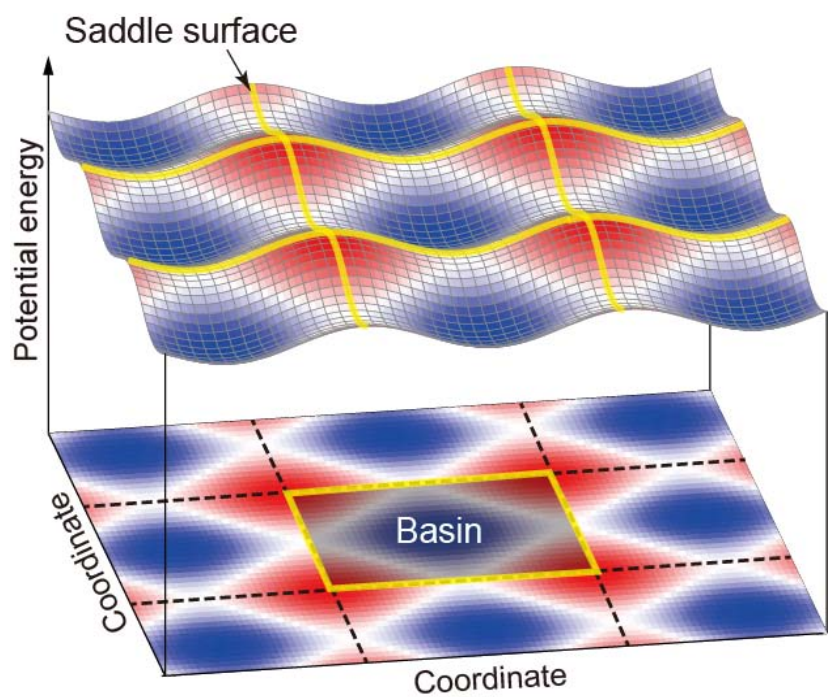

FIG. 1. (Color online) A basin surrounded by the four saddle surfaces (yellow lines) on a synthetic 2D-PES.

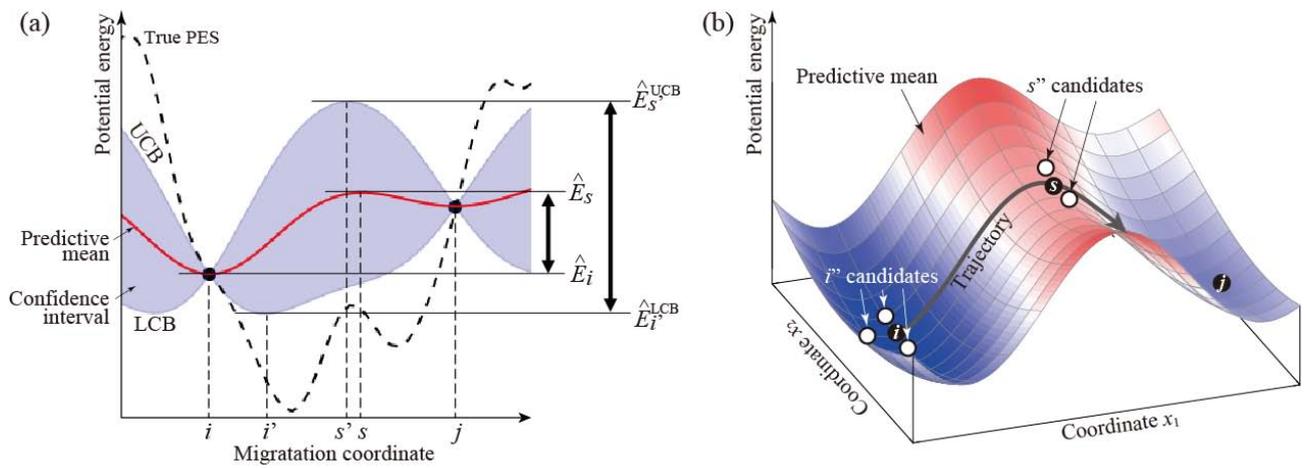

FIG. 2. (Color online) (a) Schematic diagram of the GP-PES along the trajectory of an elementary process from the initial point $i$ to the final point $j$ through the saddle point $s$. The red line and the pale blue area denote the predictive mean and the confidence interval, respectively. (b) Candidates of points $i$" and $s$" on the predictive mean of the GP-PES (two-dimensional PES), which are adjacent to the initial and saddle points, respectively. Note that the grid points on the trajectory of the elementary process are excluded from the $i$" and $s$" candidates.

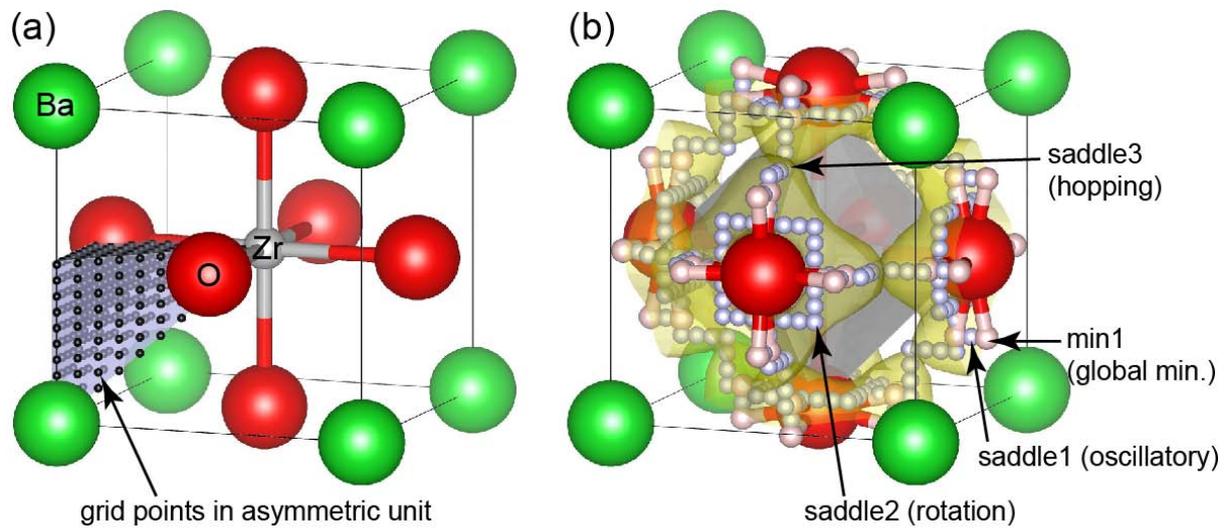

FIG. 3. (Color online) (a) All irreducible grid points in the asymmetric unit of the $c$-BaZrO$_3$ crystal. (b) The global minimum points (pale red spheres) and the trajectories (pale blue spheres) of three types of elementary processes for proton jumps in $c$-BaZrO$_3$ (*oscillatory proton transfer*, *proton rotation*, and *proton hopping*). The yellow surface denotes the PE isosurface (isosurface level: 0.3 eV vs. the global minimum point).

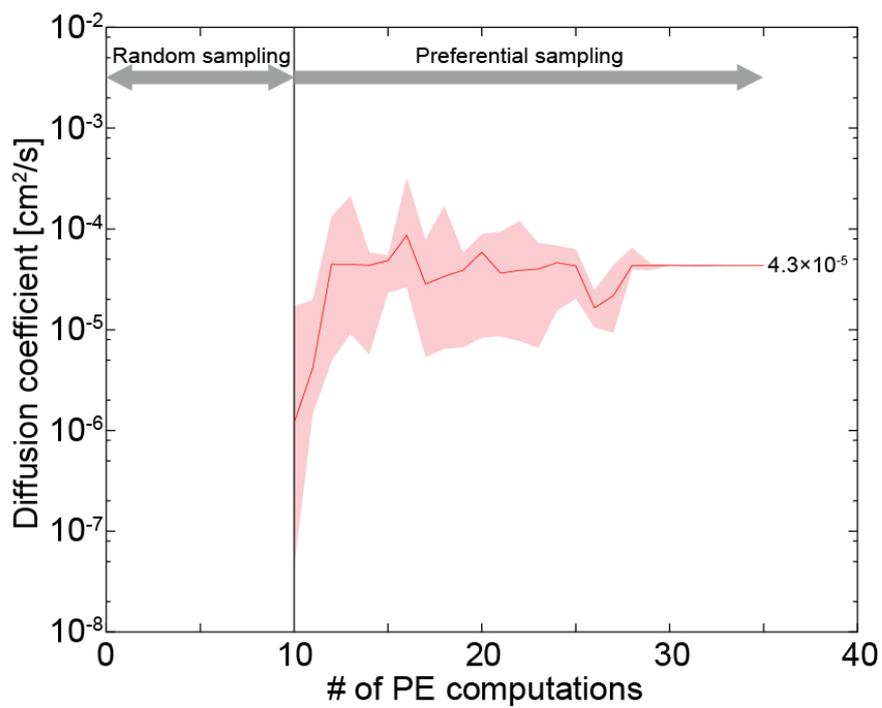

FIG. 4. (Color online) The predicted diffusion coefficient of protons at 1000 K in $c$-BaZrO$_3$ with the uncertainty as a function of the number of PE computations. The first ten PE computations corresponds to the initial points sampled at random. The red line is the diffusion coefficient on the predictive mean of the GP-PES. The pale red region denotes the uncertainty of the predicted diffusion coefficient estimated from the 95 % confidence interval of the GP-PES.

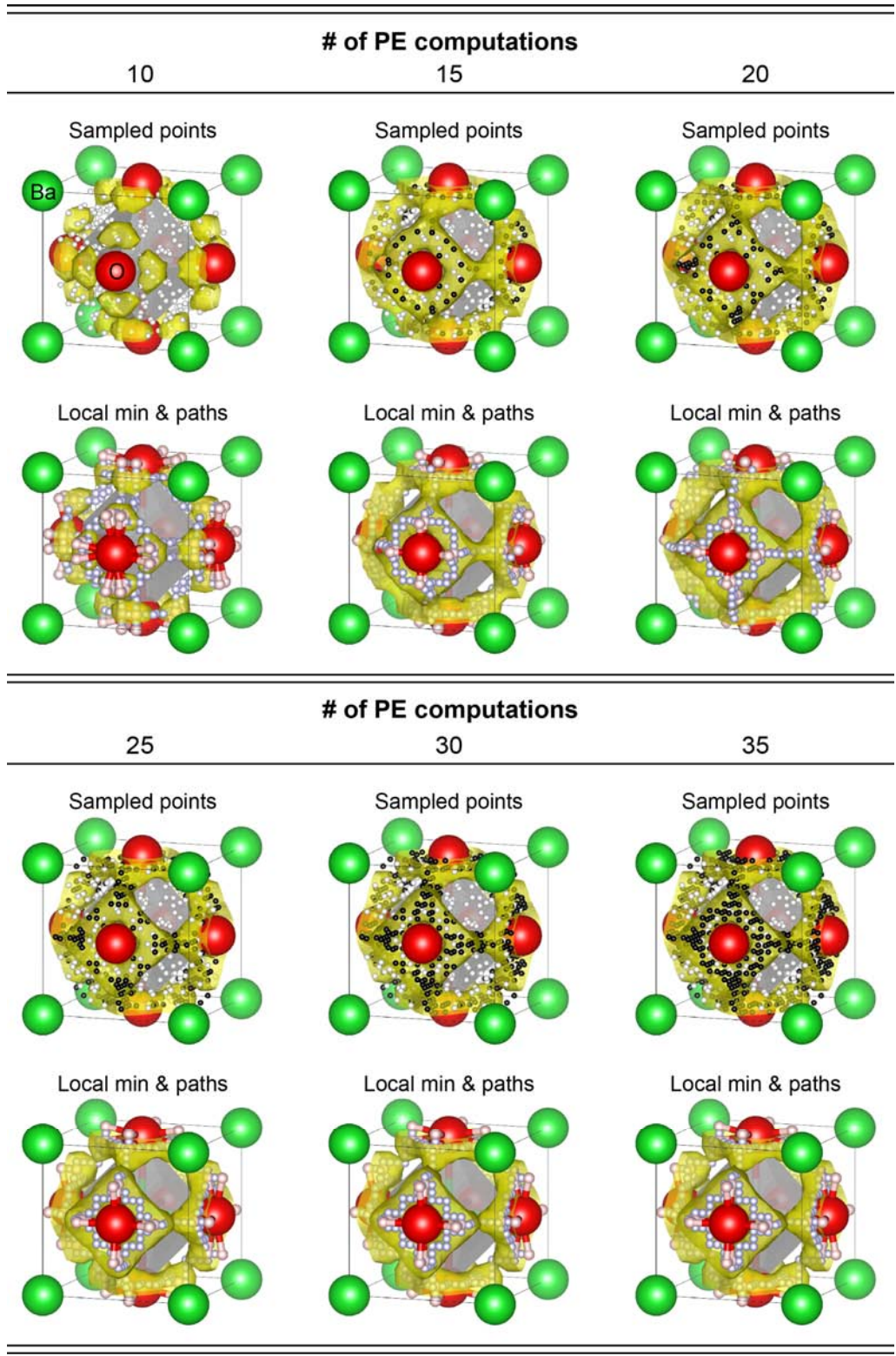

FIG. 5. (Color online) The sampled grid points (white and black spheres), local minima (pale red spheres) and trajectories of elementary processes for proton jumps (pale blue spheres) on the predictive mean of the GP-PES after 10, 15, 20, 25, 30, and 35 PE computations. The white and black spheres correspond to the initial random sampling and the subsequent preferential sampling, respectively. The yellow surface denotes the PE isosurface on the predictive mean of the GP-PES (isosurface level: 0.3 eV vs. the global minimum point).

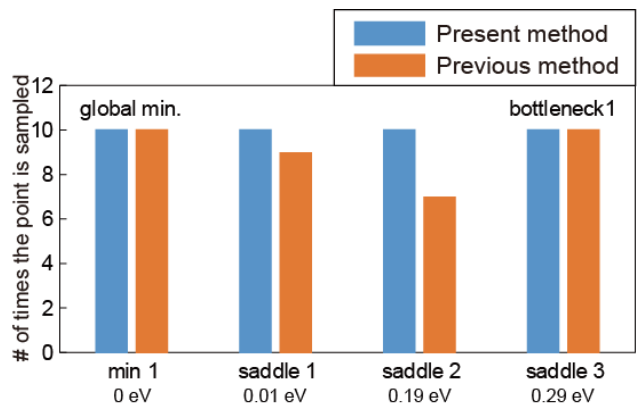

FIG. 6. (Color online) The comparison between the present method and the previous method [10] in the number of times that the global minimum and three saddle points in $c$-BaZrO$_3$ were sampled in the ten trials.

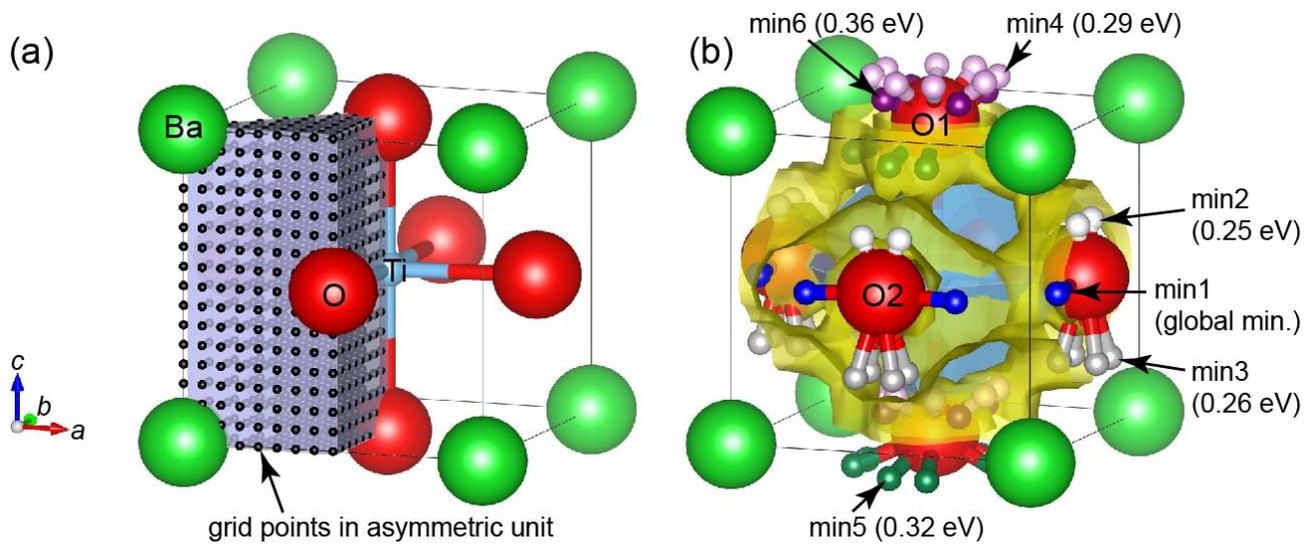

FIG. 7. (Color online) (a) All irreducible grid points in the asymmetric unit of the *t*-BaTiO$_3$ crystal. (b) The six local energy minima (small spheres bonding to single oxide ions) in *t*-BaTiO$_3$. The relative potential energies at the local energy minima are shown in parentheses. The yellow surface denotes the PE isosurface (isosurface level: 0.5 eV vs. the global minimum point).

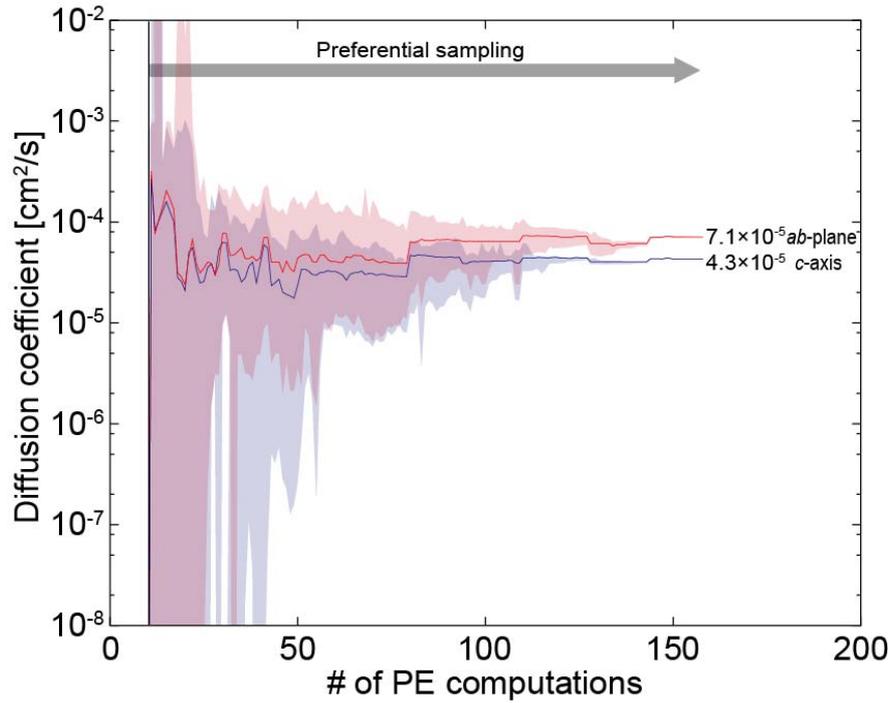

FIG. 8. (Color online) The predicted diffusion coefficients of protons at 1000 K in $t$-BaTiO$_3$ with the uncertainty as a function of the number of PE computations. The first ten PE computations correspond to the initial points sampled at random. The red and blue lines are the diffusion coefficients in the $ab$-plane and along the $c$-axis on the predictive mean of the GP-PES, respectively. The pale red and blue regions denote the uncertainties of the predicted diffusion coefficients estimated from the 95 % confidence interval of the GP-PES.

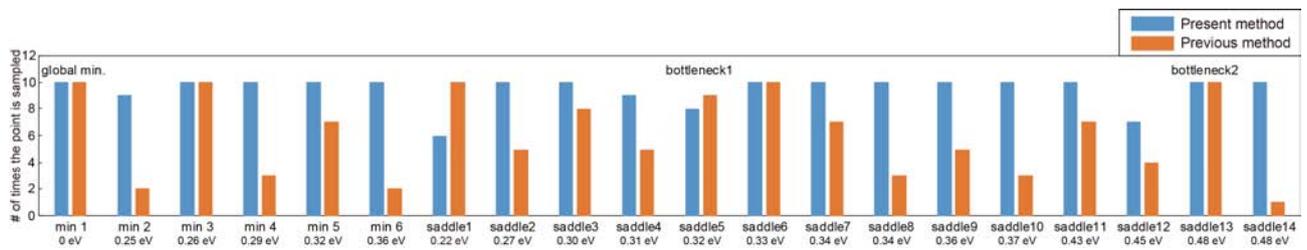

FIG. 9. (Color online) The comparison between the present method and the previous method [10] in the number of times that the true local minima and saddle points in $t$-BaTiO$_3$ were sampled in the ten trials.

# A Sampling Strategy in Efficient Potential Energy Surface Mapping for Predicting Atomic Diffusivity in Crystals by Machine Learning

Kazuaki Toyoura[*,1,2], Takeo Fujii[1], Kenta Kanamori[3], and Ichiro Takeuchi[2,3]

[1]Department of Materials Science and Engineering, Kyoto University, Kyoto 606-8501, Japan
[2]RIKEN Center for Advanced Intelligence Project, Tokyo 103-0027, Japan
[3]Department of Computer Science, Nagoya Institute of Technology, Nagoya 466-8555, Japan

The proposed method in the present study has been developed for preferential potential energy surface (PES) mapping to efficiently predict atomic diffusivity in crystals. This method is applied to two examples, i.e., the isotropic and anisotropic proton diffusion in $c$-BaZrO$_3$ and $t$-BaTiO$_3$ with the cubic and tetragonal perovskite structures. Ten trials with different initial grid points were performed for each of the example systems.

Figures S1(a) shows the profiles of the predicted diffusion coefficient of protons in $c$-BaZrO$_3$ in the ten trials. For comparison, Fig. S1(b) shows the profiles of the diffusion coefficient estimated on the predictive mean of GP-PES in our previous method*.

Figures S2 and S3 show the profiles in the case of the anisotropic proton diffusion in $t$-BaTiO$_3$, which are characterized by two diffusion coefficients in the $ab$-plane and along the $c$-axis. (a)(b) and (c)(d) correspond to the proton diffusion in the $ab$-plane and along the $c$-axis, respectively. (a)(c) and (b)(d) are the results by the present method and the previous method*, respectively.

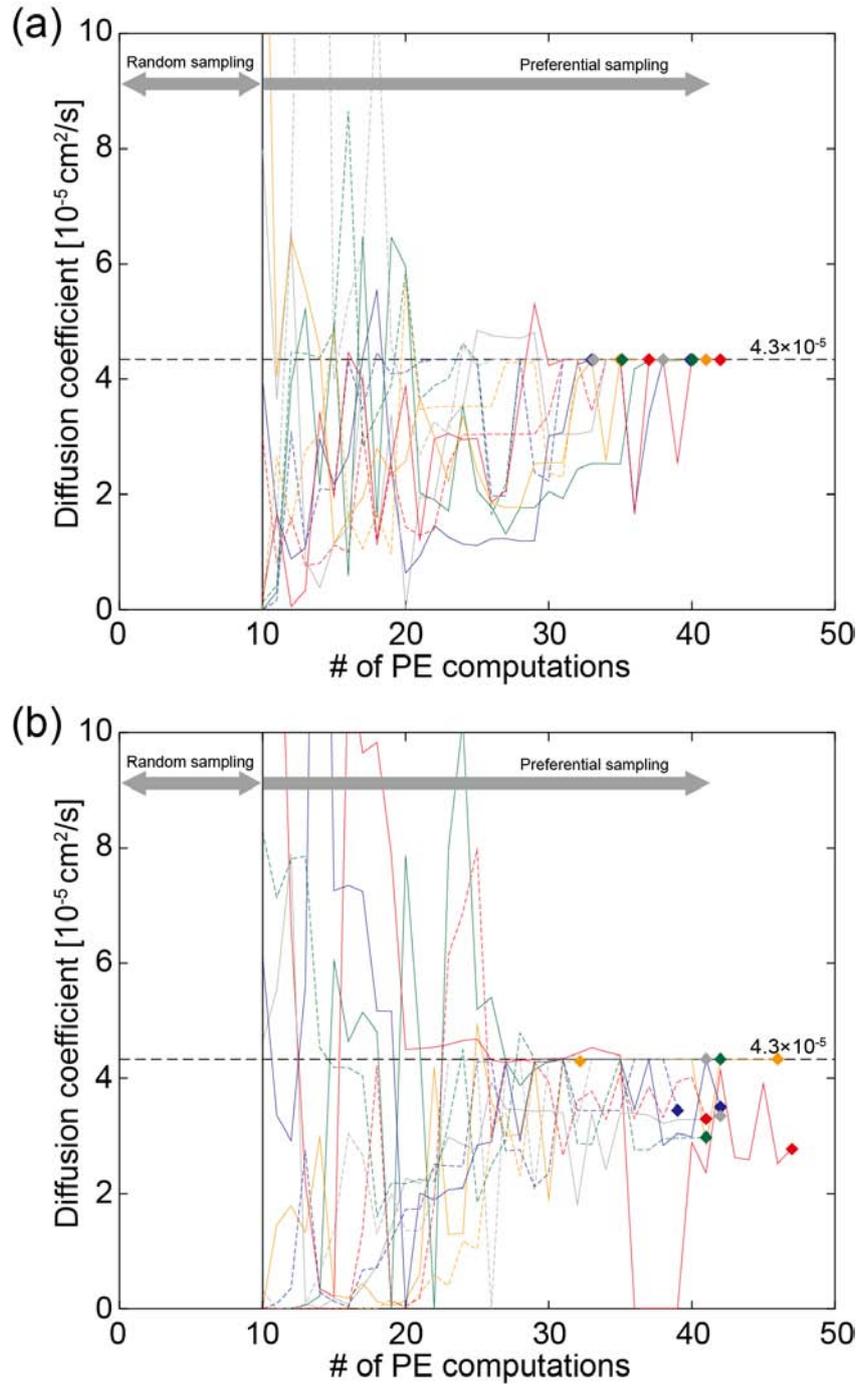

FIG. S1. (Color online) Profiles of the predicted diffusion coefficient of protons in ten trials with different initial grid points in the case of the isotropic proton diffusion in $c$-BaZrO$_3$. (a) and (b) show the profiles in the present method and the previous method*, respectively. The solid symbols denote the final values at these trials.

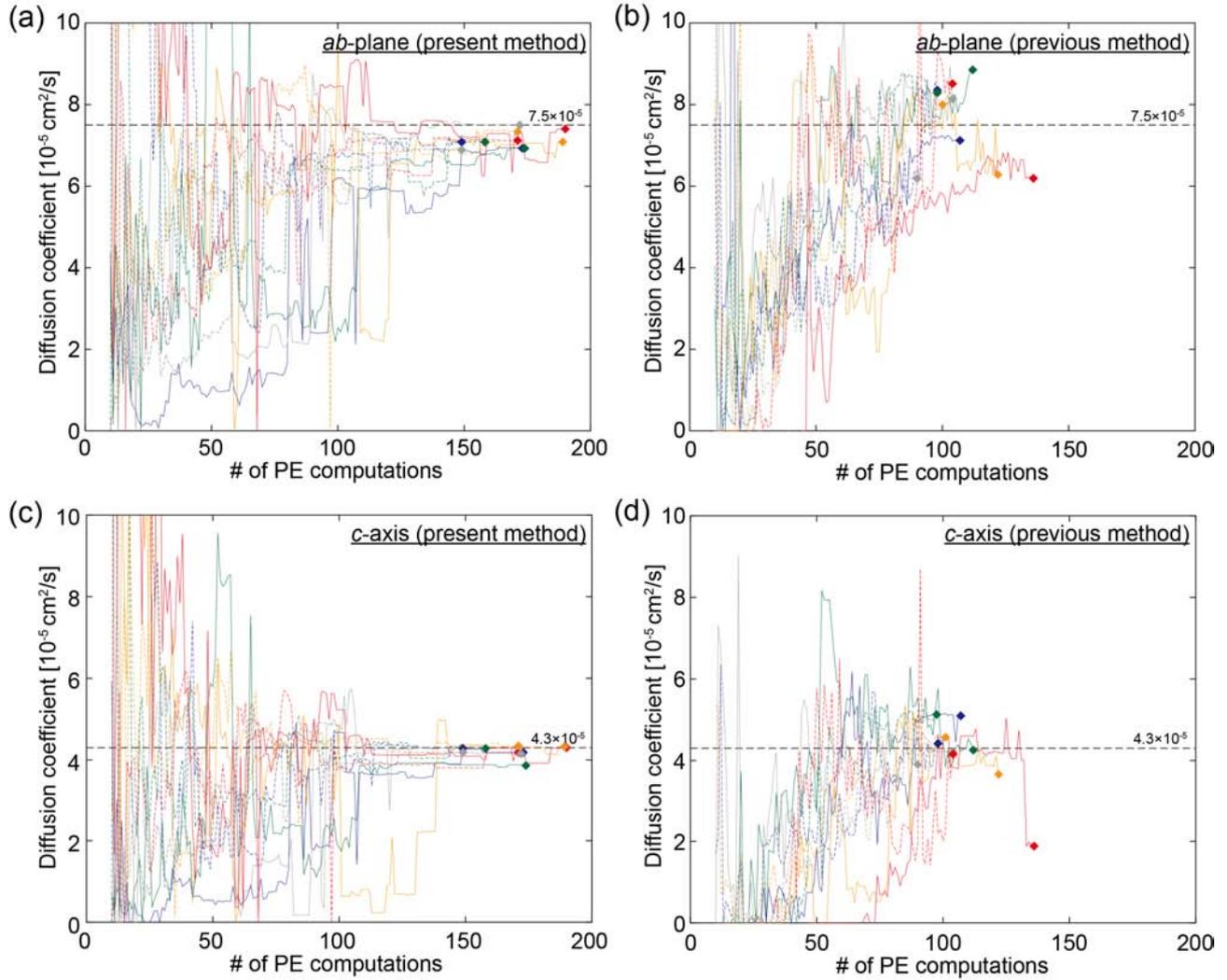

FIG. S2. (Color online) Profiles of the predicted diffusion coefficients of protons in ten trials with different initial grid points in the case of the anisotropic proton diffusion in $t$-BaTiO$_3$. (a)(b) and (c)(d) correspond to the diffusion coefficients in the $ab$-plane and along the $c$-axis, respectively. (a)(c) and (b)(d) show the profiles in this method and the previous method*, respectively. The solid symbols denote the final values at these trials.

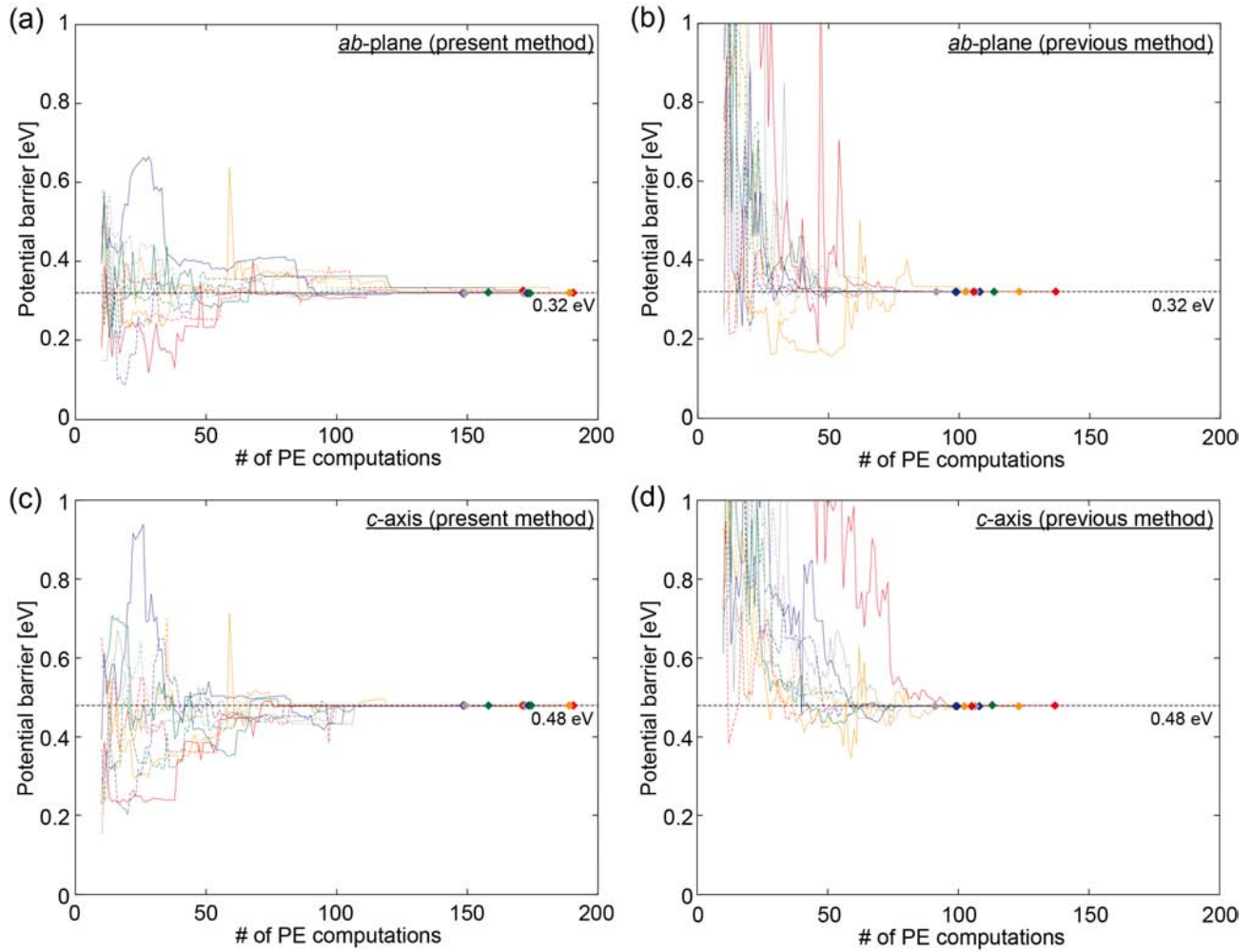

FIG. S3. (Color online) Profiles of the predicted potential barriers of ten trials with different initial grid points in the case of the anisotropic proton diffusion in $t$-BaTiO$_3$. (a)(b) and (c)(d) correspond to the potential barriers in the $ab$-plane and along the $c$-axis, respectively. (a)(c) and (b)(d) show the profiles in this method and the previous method*, respectively. The solid symbols denote the final values at these trials.